\documentclass{llncs}
\usepackage{makeidx}  
\usepackage{graphicx}
\usepackage{subfig}
\usepackage{amsmath}
\begin{document}
\mainmatter              
\title{The Formulator MathML Editor Project:\\
User-Friendly Authoring of Content Markup Documents}
\author{Andriy Kovalchuk \and Vyacheslav Levitsky \and
Igor Samolyuk \and Valentyn Yanchuk}
\institute{Zhytomyr State Technological University,\\
Chernyakhivskogo 103, 10005 Zhytomyr, Ukraine,\\
\email{info@mmlsoft.com},\\
\texttt{http://www.mmlsoft.com}
}

\maketitle              

\begin{abstract}
Implementation of an editing process for Content MathML formulas in common visual style is a real challenge for a software developer who does not really want the user to have to understand the structure of Content MathML in order to edit an expression, since it is expected that users are often not that technically minded. In this paper, we demonstrate how this aim is achieved in the context of the Formulator project and discuss features of this MathML editor, which provides a user with a WYSIWYG editing style while authoring MathML documents with Content or mixed markup. We also present the approach taken to enhance availability of the MathML editor to end-users, demonstrating an online version of the editor that runs inside a Web browser.
\footnotetext[1]{The final publication of this paper is available at \texttt{www.springerlink.com}}
\end{abstract}
\begin{keywordname}
Content MathML, mathematical formula, natural editing of algebraic expressions, model-based editor, online MathML equation editor
\end{keywordname}
\section{Introduction}
Modern standards for representations of mathematical knowledge and easily accessible software tools greatly benefit education, and scientific and technical publishing. For instance, a good number of software systems supporting the MathML standard have been helping to develop valuable distance learning models, web-based education, and electronic textbooks, which would not be available if one could only use static images, prepared in advance, instead of more natural ways for students and teachers to present and exchange formulas and diagrams. Such communication including mathematical data fosters an understanding between teachers and students when studying disciplines with strong mathematical backgrounds, since it is necessary not only  to illustrate ideas, but also to have confirmation that the student has gained insight into the material.

A rendering of mathematical notation is not the only thing needed in such scenarios, because it is the underlying mathematical structure of an expression that must be considered (and maybe examined using additional software), and not any particular rendering of an expression. This means that, in the face of poor software support for mathematical content standards, this communication process takes a turn for the worse. There are good ways to represent and reuse mathematical content data with existing standards (the content markup part of MathML \cite{1}, OpenMath \cite{2}), and so it is becoming increasingly important to find ways of expanding our ability to support these standards in software.

This paper presents a short description of one such software system, namely the Formulator MathML Editor Project \cite{3} (\texttt{http://www.mmlsoft.com}). We demonstrate how support for a mathematical content standard is achieved in the context of Formulator project and discuss features of this MathML editor that provides a user with a WYSIWYG editing style while authoring MathML documents with content or mixed markup.
\section{Formulator MathML Editor}
\subsection{Challenges of Content MathML support}
A lot of software systems for authoring mathematical documents are available, and many of them have support for MathML, either directly or through some conversion. Many references to well-known software tools and research or prototype projects can be found on the MathML Software list maintained by W3C \cite{4}. However, if we have in mind support of mathematical content standards, it does not matter how many good products help one to type in a mathematical expression if they cannot help produce some mathematical content output.

Filtered by requiring support for Content MathML, the list still includes many interesting projects, e.g., the semantically oriented formula editor WIRIS OM Tools \cite{5}, Integre MathML Equation Editor \cite{6}, MathDox \cite{10}, and new attempts to bring an interactive editing process to the web, like the Connexions MathML Editor \cite{7}.

However, an exacting user, seeking to meet the requirements of a better and quicker interaction in the field of expressing mathematical ideas, would even now not be completely satisfied for different reasons in each case \cite{8}\cite{16}, and this is natural in view of intricacy of the task. One of the important causes of failure to satisfy a user's wishes is that there is more to making mathematical information a useful resource in interactive applications than merely bringing together a common set of components and attributes adopted from the standard. Visual representation and user friendliness of an interface are ultimately also expected by users. The main usability issue of existing software systems is that a user has to understand the structure of Content MathML in order to edit an expression, and this is a considerable disadvantage, since users are often not that technically minded.

An extensive review of common pitfalls and a comparison of behaviors for several editors for mathematical content is given in \cite{8}. Among the issues mentioned there, an important place is occupied by the problem of exposing the internal document model to a user. Presentation-oriented editors can solve the problems of edit point support and understandable navigation rules easily, but such issues can be a serious obstacle for content editors trying to be user-friendly if they consider the underlying standard of mathematical content encoding to be a central source of the editing procedure.

True user friendliness cannot be achieved using the mechanical approach of just wrapping Content MathML constructs in buttons and menu items. In such systems, a developer allows a user to forget Content MathML entity names, but still forces the user to bear in mind the tree-like structures of a text while creating a mathematical expression, not to mention a need for post-editing. 
Obviously there is a big distance between a usual mathematical editing concentrated on the visual representation of an expression, and the confusing effort to mentally synchronize rendering and semantic aspects of an expression. The last differs quite dramatically both from our experience of using legacy systems with linear mathematical input, and from the exploitation of two-dimensional input forms usual for mathematical equation editors.

These obstacles are natural consequences of attempts to represent the semantics in a different way from purely mental exercises. Any kind of user friendliness in such matters entails compromises between a shape and contents. So, in addition to general problems of human-computer interaction, concerned software developers must address a number of supplementary issues.  Demanding a well-formed and finished Content MathML document is not a great help to a developer who needs to build an interactive system. During the editing process, there are ambiguities which either can only be solved at the cost of a deeper analysis of the context, or are doomed to be left unresolved pending user hints. The first approach of context analysis leads to performance problems, the second is not user friendly.  But to neglect resolving ambiguities is also not a good way if one is trying to provide lasting correctness of the mathematical content output.

Thus we have a classical triangle of alternatives, where software performance, permanent output correctness, and user friendliness occupy three corners, and existing systems far too often implement only a single edge in this triangle. While
having every respect for the efforts and achievements of our colleagues in the field and by no means pretending to cover in full the triangle of alternatives for Content MathML editing, we hopefully have some advances in the Formulator MathML Editor Project which helps to bring a process of authoring MathML documents with content or mixed markup nearer to a user's needs.
\subsection{Overview of the Formulator MathML Editor}
The Formulator MathML Editor Project was started in 2003 as part of an on-site computer algebra system \cite{11}. Since than the project has grown from a product-oriented mathematical expression editor to a set of desktop and online software tools for editing and rendering MathML documents. Over the years, a number of successful use cases have shown the suitability of the Formulator software for a wide audience of software developers, educators, authors, and students. Among featured projects are the Standards Unit Mathematics Project at the UK DfES \cite{13}, commercial applications, regional education service centers and educational institutions in the USA and EU.

The main components of the Formulator software are:

\begin{itemize}
\item {\itshape Formulator MathML Weaver}: a desktop application for  WYSIWYG editing of Presentation and Content MathML documents. There is a proprietary version for MS Windows and an open source cross-platform version, available from SourceForge and Google Code sites.
\\
\item {\itshape Mathematical Templates Builder}: an utility for customizing and amending a dictionary of mathematical symbols and templates. This tool ensures Formulator openness in the sense that a user can change the look and style of a set of existing mathematical constructs without accessing the software source code. This is possible because templates are explicitly defined and can be edited in text form, and a simple built-in language is available to specify the dynamic behavior of graphics and to edit input slots in some complicated situations. During a run-time the built-in language provides a way to calculate coordinates and sizes of graphics objects and edit boxes, margins of a template and its child objects. The simple structure of the language guarantees its fast run-time interpretation and leads to effective rendering of equations.
\\
\item {\itshape Formulator ActiveX Control and API}: component editions of the desktop editor which are intended for a software developer who needs to build an application aware of the mathematical typesetting and semantic rules.
\\
\item {\itshape Formulator IE Performer}: a plug-in for Internet Explorer to render MathML fragments inside web pages (similar to MathPlayer \cite{12}).
\\
\item {\itshape Online MathML Editor}: a MathML editor in the form of a distributed web application that should be run inside an Internet browser.
\end{itemize}

Considered in the role of a presentation formula editor Formulator is similar to a majority of mathematical expression editors embedded in office products such as OpenOffice.org and MS Office, and to well-known WYSIWYG formula editors, for example, MathType \cite{12}. Among its notable supplements are several dedicated model views for clearer understanding and fine tuning of the MathML document structure, and additional tools for developers and advanced users to enable enhancing the editor's functionality and to build new software applications which use the formula editor.

As a content editing system the Formulator software is far from being completely satisfactory yet, and we definitely lack feedback from existing users worldwide and large-scale user studies. On the other hand, a frame of reference can be achieved using comparison with content editors mentioned earlier, from the methodological and survey papers such as \cite{8}\cite{9}, from feedback about Formulator's suitability for learners (the Autograph Maths software \cite{15}) and instructors (the Connexions project \cite{14}).

As a mathematical content editor the Formulator project is designed from the start as an intuitive and visually oriented tool. It offers what the paper  \cite{9} calls ``natural editing of algebraic expressions'' with the operations, natural for a presentation formula editor, of insertion, deletion, selection, cut, copy, paste, drag and drop.

The modes of editing and navigation are consistent with the ``what-you-see-is-what-you-get'' principle, supporting edit-point feedback and accessibility, geometric moves, reversing of moves, deterministic moves, slot navigation and selection flexibility \cite{8}. In addition to simple ``natural editing'' procedures, the Formulator editor has support for advanced modes, which are described in \cite{9} for the Aplusix software and WIRIS OM Tools: syntactically basic enhancements (recognition of mathematical operators and parentheses, arity representation), enhanced backspace and delete operations, algebraic selections, and automatic bracketing in accordance with changing operator precedence.

The model chosen for the Formulator editor to enhance backspace, delete and insert operations differs from that proposed in \cite{9}. The latter approach has an advantage in the case of unary operators and derives a deeper benefit from insights into user intentions, for instance, in manipulations with selections and operand movements. The Formulator MathML editor employs semantic information about a document in a different way, for example, by evaluating simple Content MathML formulas and formula chains with user variables. We expect to enhance the editing process in the Formulator editor further by additional utilization of semantic information.
\subsection{General Approach to the Document Structure}
The nature of Content MathML differs radically from Presentation MathML, as they are used to define different aspects of mathematics. Presentation elements describe the visually important two-dimensional hierarchical forms and thus give more or less precise instructions on how to render and how to edit mathematical constructs. Content markup follows closely the semantic structure of mathematical objects. Consolidation of these two quite different formats into one document structure seems overly complicated from a software implementation point of view, and it appears a good idea to have one dominant format.

Since initially the Formulator MathML editor was as a tool for Presentation MathML authoring, our aim was to bring semantics into existing visual editing and to retain user friendliness. In this way, the document structure in Formulator has support for Presentation MathML, both encouraging development of the editing process of Content MathML formulas in a common visual style without exposing implementation details of the Content MathML format, and keeping good software performance while rendering and editing the document. The internal document model is a tree of four basic node types, holding additional attributes as reference information: (1) an input slot, (2) a line of horizontally neighboring text and formula nodes, (3) a formula, composed of input slots and graphics, and (4) plain text.

This document structure allows carrying out the initial task of creating Content MathML expressions by relating each semantic construct with corresponding visual elements and by adding supplementary nodes, which are invisible either from rendering or from the semantic point of view. Further free-style operations also require more labeling throughout a document to account for intermediate editing states, which inevitably break the proper Content MathML format. Thus a compromise must be found between maintaining semantic correctness of the document model and user operations which are incompatible with expression semantics (e.g., brackets which are not balanced).

For instance, in the context of Presentation MathML the formula ``\(2 \times 3 + 4\)'' is presented by the trivial hierarchy of fig.~\ref{fig:1a}, but an essentially more verbose document structure is needed to implement the free style of editing in the context of Content MathML, fig.~\ref{fig:1b}.

\newsavebox{\tempbox}
\sbox{\tempbox}{\includegraphics[scale=0.5]{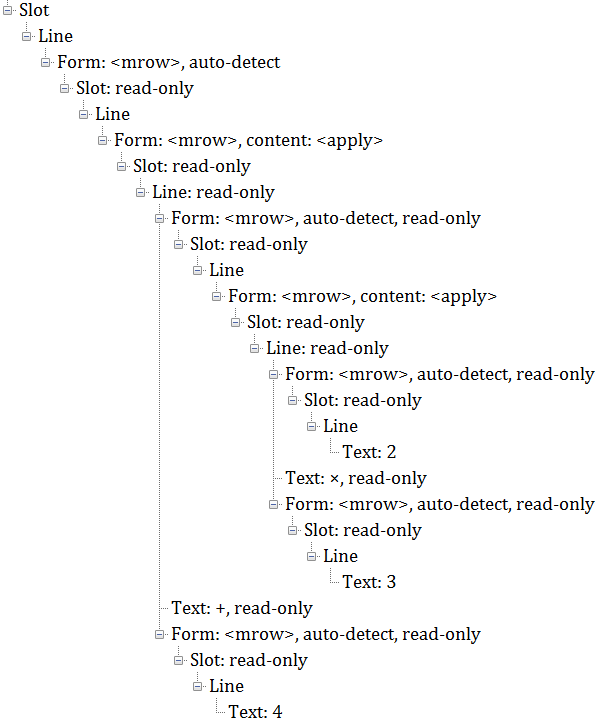}}%
\begin{figure}
  \centering
  \quad
  \subfloat[]{\label{fig:1a}\vbox to \ht\tempbox{\vfil\hbox to 40pt{\includegraphics[scale=0.5]{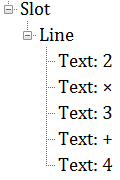}}\vfil}}
  \qquad \qquad
  \subfloat[]{\label{fig:1b}\includegraphics[scale=0.5]{./fig1b.png}}
  \quad
  \caption[]{Document structure for a formula ``$2 \times 3 + 4$'' in the context of Presentation MathML (a) and Content MathML (b). After each formula node, there is an indication of a Presentation MathML tag that is used by a rendering procedure and either a type of the node in the Content MathML context, or an ``auto-detect'' node attribute. Read-only nodes cannot be either deleted or moved by a user.}
  \label{fig:1}
\end{figure}

In addition to more complicated document structure, there are attributes and hidden reference nodes, which guarantee correctness of the formula during editing actions and are used to produce proper Content MathML output from this structure. E.g., ``do not edit'' and ``do not move'' attributes are used to preserve the mathematical template and to provide a user with comfortable navigation through nodes (no ``fake moves'' \cite{8}, each user action corresponds to an understandable movement of the caret marking the insertion point). Some formula nodes are marked with ``auto-detect self entity by contents'' attributes and do not recognize their type until a user finishes editing and wants to save or to export a formula to MathML. The combination of these attributes allows one to remove parts of a formula and to insert quite different kinds of elements, for instance, a division operator instead of a token element, and still to have a proper output. The ``auto-detect'' attribute is one of the reasons for the verbosity in a document structure, but at the same time it is a way to avoid the rigidity of a template-based editing procedure.
\subsection{Improving Usability for a Template-Based Approach}
Earlier versions of the Formulator MathML editor implemented a template-based approach to Content MathML creation, where each new construct was chosen from a palette of atomic formulas and had to be placed into some input slot of previously used constructs.

Users in experiments we conducted with creating and post-editing of content formulas gave us feedback on usability issues, concentrating on the  uncomfortable rigidity of templates with a read-only structure and several edit spots, and on the confusing view of the editing process as building a tree. On the other hand, just considering this approach to be faulty seems to be going too far. Amongst the good ideas of a template-based approach are supporting a user with run-time information about proper formula structures, quick construction of new formulas from basic building blocks, and an editing style that facilitates general correctness of a document from a semantic point of view.

In moving towards improved usability of Content MathML support, we prefer evolutionary changes rather than revolutionary ones, and so combine the existing positive features of the template-based approach with a newly implemented free style of editing. The change is to replace rigid templates with a more fluid structure that can be changed by a user, instead of the former frustrating experience of having to delete a template as a whole.

An important aspect of this proposed free-style editing is that changing the structure of a Content MathML formula is closer to the  behavior of a state machine than, for instance, to plain text editing. This means that we preserve run-time hints and explicit correctness of the template-based approach by transforming one proper structure of an expression to another one as a response to user insertions and deletions.

Such a process seems more reliable and predictable in comparison with encouraging a user to go ahead and to break down an expression, with the hope that later the user will manage to reassemble the expression in a semantically proper way. The latter would be more like a computer programming, and maybe that is too much to expect from not so technically minded users.
\subsection{Examples of Free-Style Editing}
In order to realize our conception of free-style editing, we paid attention to such important aspects of the input system as random post-editing, insertion, deletion of mathematical operators and whole expressions, automatic completion of an expression, and a model for bracket editing.

\begin{figure}
  \centering
  \subfloat[]{\label{fig:2a}\includegraphics[scale=0.50]{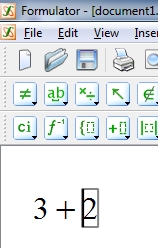}}
  \hspace{0.5cm}
  \subfloat[]{\label{fig:2b}\includegraphics[scale=0.50]{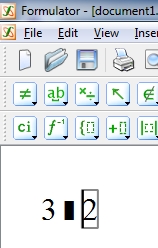}}
  \hspace{0.5cm}
  \subfloat[]{\label{fig:2c}\includegraphics[scale=0.50]{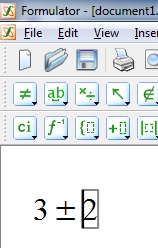}}
  \caption[]{These snapshots demonstrate how a formula can be edited without breaking the underlying template. Transition from case (a) to case (b) occurs after Backspace is pressed, and after typing `$+$' and `$-$' in sequence the case (c) appears. The latter transformation also demonstrates a feature of automatic replacement.}
  \label{fig:2}
\end{figure}

For instance, once an operator has been entered it was impossible in the earlier versions to change or delete it without deleting the entire sub-tree containing it. If a user entered ``$3+2$'' and then wanted to change it to ``$3-2$'', it was impossible to position the cursor at the `$+$' operator to change it to a `$-$'. Instead the whole template of ``$3+2$'' had to be deleted and re-entered. Now a user can edit initially read-only mathematical operators by deleting them (a black box appears instead, standing for ``no operation'', see fig.~\ref{fig:2b}), and later typing in a new mathematical operator. This black box preserves the correct expression structure, hints that the editing process should be continued to recover expression correctness, and allows a user to change it into a needed mathematical operator from a palette, while still being protected from unconditional deletion from the expression.

This example shows also that a sequence of characters can be converted (where it is possible) to a mathematical operator that has no direct equivalent on a keyboard. For instance, by typing the `$<$' operator and then immediately typing the `$=$' operator, we will get a new `$\leq$' operator. The feature of automatic replacement does not exclude any subset of proper expressions, since it can be undone if not wanted, thus leading to more complicated structures with nested mathematical operations.

Another example of free-style editing is changing the structure of a formula by explicit insertion or deletion of brackets (in contrast to the implicit brackets used where required by a combination of the existing structure of an expression and mathematical precedence information). What we are looking for is an editing procedure that is as similar to linear text entry as possible, both for the initial input of an expression and the subsequent editing of it. Again, since we do not want the user to have to understand the structure of Content MathML, the main challenges are:

\begin{itemize}
	\item to save valid MathML even when a user opens brackets and does not close them, and 
	\item to find a proper transition from an old formula structure to a new one after a selection is made that should be inserted into brackets.
\end{itemize}

\begin{figure}
  \centering
  \subfloat[]{\label{fig:3a}\includegraphics[scale=0.40]{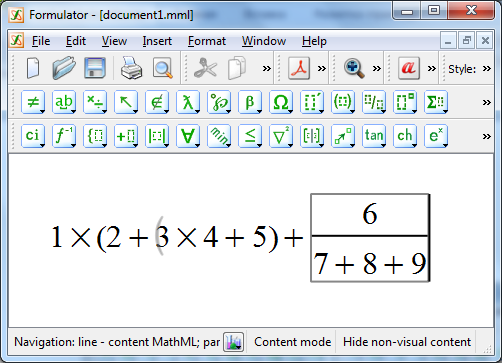}}
  \hspace{0.5cm}
  \subfloat[]{\label{fig:3b}\includegraphics[scale=0.40]{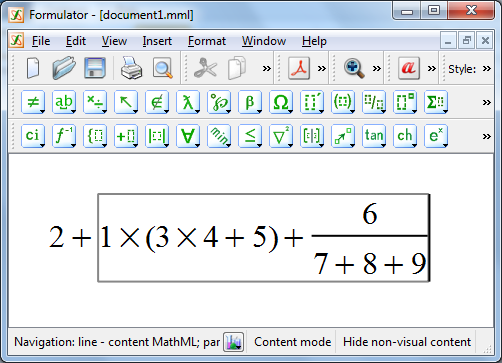}}
  \caption[]{These snapshots demonstrate semi-transparent Content MathML rendering of brackets and highlights some possible ambiguities in arbitrary bracketing. Case (a) has an unbalanced bracket to the left of number 3, and by identifying the left edge of a newly created bracketed expression to be inside existing brackets and the right edge to be on the rightmost end of the whole expression, a user splits the former expression with a result, case (b), that at first is not evident, but in a sense is logical.}
  \label{fig:3}
\end{figure}

The second problem (i.e., genuine changing of a formula structure) seems to have no universal solution because of possible ambiguities. Thus, a user can break an existing bracketed expression by enclosing only a part of this expression in a new pair of brackets along with the rest of the formula that initially lies outside the previously existing brackets (fig.~\ref{fig:3b}). Nevertheless, Formulator MathML editor implements a smart method for solving such ambiguities and where it is possible (in most cases, as our practice shows) a formula structure is changed predictably and clearly for a user.

\begin{figure}
  \centering
  \subfloat[]{\label{fig:4a}\includegraphics[scale=0.40]{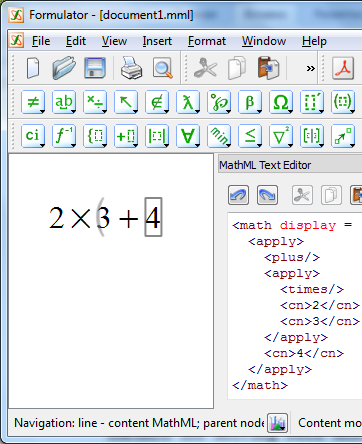}}
  \hspace{0.05cm}
  \subfloat[]{\label{fig:4b}\includegraphics[scale=0.40]{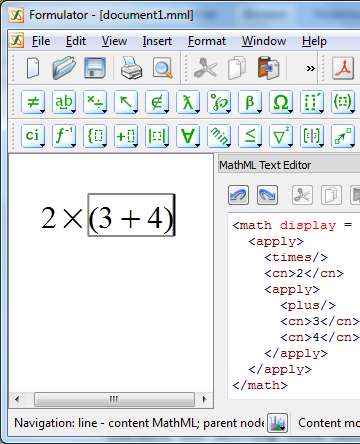}}
  \hspace{0.05cm}
  \subfloat[]{\label{fig:4c}\includegraphics[scale=0.40]{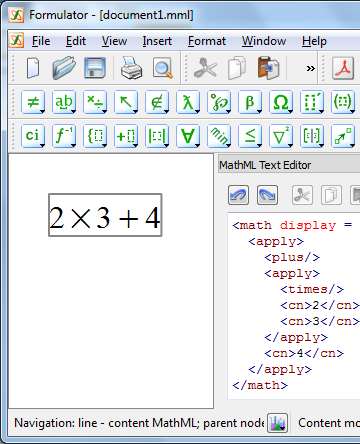}}
  \caption[]{These snapshots demonstrate how structural changes in bracketed formula can be reverted using deletion. Transition from the case (a) to (b) occurs after `$)$' is pressed at the right of the formula, and if a user presses Backspace when a caret is in this position, the case (c) appears. The latter snapshot corresponds to the structure of case (a) as we can see from plain text of Content MathML output.}
  \label{fig:4}
\end{figure}

The transitions of a formula's structure discussed above can be reverted by using further editing operations. Although a bracketed expression is treated in accordance with the underlying template as having read-only brackets and several input slots inside them, this structure is not rigid and can be altered if a user presses the Delete button and a caret is standing before the left bracket, or if a user presses the Backspace button and a caret is standing after the right bracket (fig.~\ref{fig:4}).

\subsection{Supporting Users Accustomed to Legacy Input}
There is one more enhancement of the Content MathML editing process in the Formulator MathML Editor project that can be valuable for users accustomed to legacy systems with a plain text mathematical input. This means a procedure similar to linear input that tries to express mathematical equations using just keyboard characters, for instance, the circumflex `\textasciicircum' sign for an exponentiation and the `/' sign for a fraction or division, as in ``$y=1/x \text{\textasciicircum} 2$''. Support for this input mode allows an easy transition to the Formulator entry system for users who are used to the legacy style, and also allows faster and easier input in some special cases using only a keyboard without a mouse.

Improved usability of this sort means that users are able to enter expressions intuitively with a sequence of key presses. For instance, the equation $y=\sin x$ should be entered by pressing `y', `=', `s', `i', `n', `x', and an equation $y=2a^2b$ should be entered by pressing: `y', `=', `2', `a', `a', `b'. No mathematical operator buttons or screen forms are used in either example, but as a trade-off for using such shortcuts we assume that a user has only variables of one-letter length. In addition to the legacy input style, there are also a few special cases, where a certain sequence of key presses can generate slightly different but more intuitive input strings (for instance, several sequential presses of the letter `a'  generate exponentiation instead of multiplication).

More is expected, and thus user friendly behavior is also provided by using information about operator precedence to automatically help a user to navigate through the text; namely, the editor automatically shifts a caret forward in situations where a user of a legacy software would expect this. For instance, when a user types `y', `=', `1', `/', `x', `+, `1', in basic editing mode the formula would be considered as ``$y = 1/(x+1)$'', since a caret is near the `$x$' position and the `$+$' operator is normally associated with a caret focus. In order to meet user expectations this rule is adjusted during legacy style input (by considering precedences of division and addition operators), so that we get exactly the  formula ``$y = 1/x+1$'', as a user would expect who is used to textual mathematical input.
\section{Availability and Future Work}
Guided by a vision of public accessibility of technologies for learning activities creating and authoring documents containing mathematical data, the Formulator MathML Editor project is evolving to enhance its availability to different kinds of end-users, from students to software developers.

\begin{figure}
	\centering
		\includegraphics[scale=0.33]{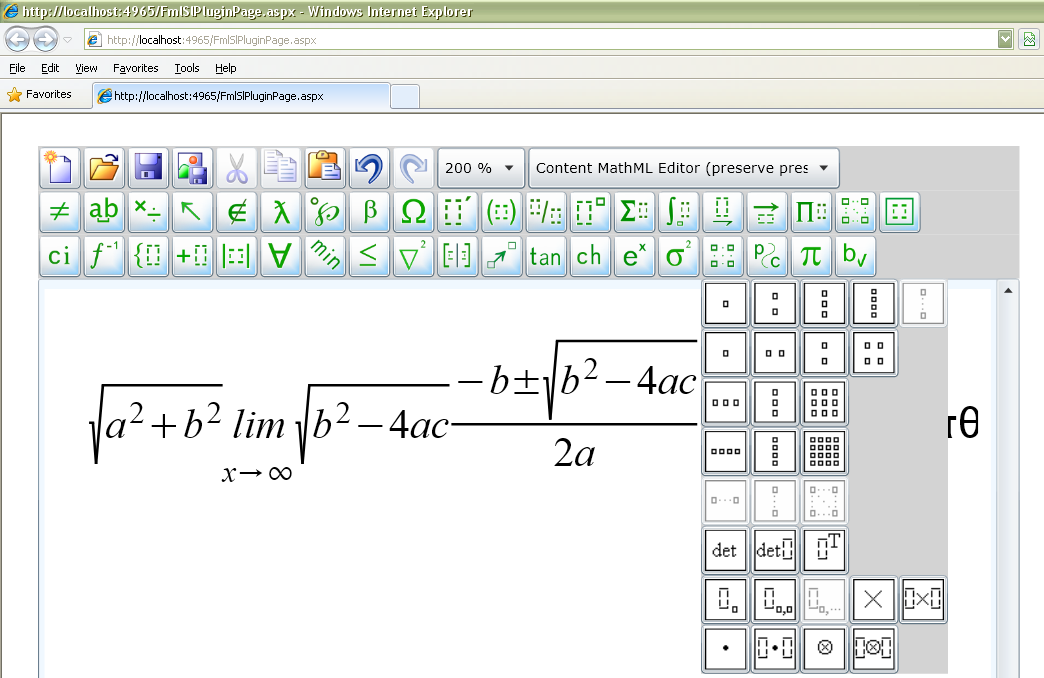}
  \caption[]{The test page of the online version of the Formulator MathML Editor. A link to the page is available from \texttt{http://www.mmlsoft.com}.}
  \label{fig:5}
\end{figure}

The main steps toward this goal have been:
\begin{enumerate}
	\item Porting the Formulator MathML Editor project to Mac OS X and Linux operating systems (in addition to MS Windows that was supported initially) and issuing this new version of the Formulator MathML Editor project with an open source license.
	\item Developing an online version of the editor that runs inside a Web browser in a form of Rich Internet Application (fig.~\ref{fig:5}).
\end{enumerate}

The second step is the more challenging, because basically the editor uses native code to get more performance and a better appearance. It turns out that a good way to solve the problem lies in using a browser plug-in (in contrast to a scripting language approach) as a client of the editor's algorithmic core that runs remotely as a native application. The general picture of this solution includes a server part and a thin Silverlight client working inside a browser;  in an ideal case this should not depend on the operating system at all. Actually, this scheme is too good to be true, and in a real world scenario the user is limited to a combination of Internet browser and operating system that already has support for the Silverlight plug-in. Furthermore, there are some delicate issues where the choice of a browser does matter (for example, when working with the Clipboard of an operating system).

It is important that generally the online Formulator editor not be restricted to the Silverlight plug-in. It should work with any other similar technology as well. The choice of the browser plug-in approach (in contrast to a scripting language method) is a more fundamental point than the choice of a Silverlight, Flash or some other plug-in.

Further improvements in the efficiency and functionality of this online version of the Formulator MathML Editor are matters for future work, although tests  which have been conducted show acceptable performance for the core editing functions already implemented. Moreover, it seems that by installing a server part within a Local Area Network of an organization where Formulator is used even more efficient behavior can be achieved. Currently, the online version of the Formulator MathML editor is available on our test server, and the set of features implemented is already sufficient to perform most popular operations, such as: to create/open/save formulas in Presentation, Content and mixed  MathML markup; to export formulas to MathML text or to an image; to cut/copy/paste formulas.

Future plans include making a thicker client, and evolving the server part to a more usable environment, for instance, closer to online word processor functionality. One more interesting opportunity that is opened after creating the Silverlight version of the Formulator MathML software is to build more complicated scenarios for education above a layer of online mathematical editing, for instance, to support interactive online learning activity.

%
%

\clearpage

\end{document}